%% file: main.tex
\documentclass[a4paper]{article}

\usepackage[utf8]{inputenc} %arXiv admin
\usepackage{INTERSPEECH2020}
\usepackage{hyperref}
\title{Aalto's End-to-End DNN systems for the INTERSPEECH 2020 Computational Paralinguistics Challenge} % End-to-End DNN Systems for the Elderly Emotion, Breathing and Masks Challenges}
\name{Tamás Grósz, Mittul Singh, Sudarsana Reddy Kadiri, Hemant Kathania, Mikko Kurimo}
\address{
 Department of Signal Processing and Acoustics, Aalto University, Finland}
\email{firstname.lastname@aalto.fi}

\begin{document}

\maketitle
\begin{abstract}
\input{abstract}
\end{abstract}
\noindent\textbf{Index Terms}: computational paralinguistics, DNN, end-to-end, ensemble learning %multitask learning

\input{introduction}

\section{Methods}
In this section, we describe the end-to-end system usage in an ensemble learning scheme. We also present task-specific modifications to capture task requirements in the end-to-end system.  Keep in mind that this paper describes our solutions for a competition, so we broke with the tradition of using only a few techniques. Instead, we used several to get the best results. Still, we did our best to measure the impact of each modification on the development data and tested only the best ones.

\subsection{End-to-end learning}
%Basic principle
End-to-end learning is a new emerging paradigm within deep learning. Researchers across various fields have adopted this paradigm supported by the availability of large data and powerful computational resources. Theoretically,  end-to-end systems are built to replace the traditional pipeline-based solutions with a single deep neural network. The end-to-end systems allow using a single optimization step to training the complete model. They also have the promise of bypassing the laborious feature-engineering step by having a single system for solving every aspect of the prediction problem. In practice, however, these systems are built on top of existing features. The advantages of this paradigm make it an attractive choice for ComParE tasks. 

In our experiments, we employ the same DNN model architecture for elderly and mask sub-challenges. For the breathing sub-challenge, we use a different DNN architecture based on the baseline system for further research. We describe the details of these end-to-end model architectures in Section \ref{ssec:models}.
%Specific how we used this
%Lately, End-to-end DNNs are getting very popular in automatic speech recognition. They consist of a single DNN that functions as both acoustic and language models. Usually, some spectral features are feed to them, and on their output, they produce words. Based on their success, we decided to use end-to-end trained models for paralinguistic tasks, instead of using DNNs for feature extraction like auDeep and DeepSpectrum.
Our models process either spectral input features or raw audio signals in case of the breathing task. Then the DNNs can directly be optimized to perform the given task. Using this single model approach allowed us to quickly modify the general framework to the specialties of the sub-challenges.

\subsection{Ensemble learning}
%emphasize that DNNs are sensitive to random seed
DNNs are known to be sensitive to the random initialization, and our experiments also confirm this. This issue is especially severe if the amount of training data is limited, which is usually the case for paralinguistic tasks. A solution to this problem is applying ensemble learning. We train several differently initialized DNNs and then combine their predictions to get stable and even better results. 

Here, we employ a specific bootstrap aggregation method, called bagging. Originally, bagging trains each model using only a random subset of the training data to produce diverse systems. As the training data is already limited, we decide to use all available data during training and rely on random initialization and data shuffling to produce a diverse set of DNNs. In the combination, we average the outputs of differently-initialized DNN together to make the final prediction. 

The ensemble learning can also be performed with other approaches. % We can also combine different approaches in an unweighted average to leverage their individual strengths. 
For our mask sub-challenge experiments, we perform an equal-weighted soft-voting-based combination of the baseline prediction system like Support Vector Machines (SVM) with our ensemble DNNs. %In our experiments, we observed the benefits of using these combinations.

\subsection{Multiple objectives}
Training an end-to-end system does not have to be restricted to using a single loss function. Often multiple losses are taken into consideration to focus on multiple aspects of the prediction problem. This technique also helps regularize training. 

For breathing sub-challenge, the end-to-end baseline system is trained with a correlation-based loss. However, it does not help to bound the outputs to the same scale as the label. To match the output's scale to the label, we use a combination of the correlation loss and the mean squared error (MSE), which can help regularize the end-to-end baseline system.

\subsection{Multi-task learning}
Multi-task learning trains a single model to perform multiple tasks simultaneously. Recent work~\cite{Latif2020Multitask} has also shown the benefits of using this scheme for paralinguistic tasks. Intuitively, multi-task learning's unified model allows data augmentation by sharing information relevant for one task with the other. This intuition is especially relevant in the case of elderly sub-challenge. The arousal and valence levels are two related dimensions to describe the emotional experiences of the speaker. Thus, we experiment with a single end-to-end model trained to predict the arousal and valence levels in a joint framework.

\subsection{Resampling strategies for multitask learning}
\label{sec:resample}
In the elderly sub-challenge, we observe a class imbalance problem. Having over-represented classes in the data is a common problem for paralinguistic problems \cite{GosztolyaBGT17}. To address the data imbalance, we choose two sampling techniques: upsampling and probabilistic sampling~\cite{GosztolyaBGT17}. Upsampling is a simple method that repeats the underrepresented examples until the data becomes balanced. Probabilistic sampling applies a more rigorous approach. It defines the desired class distribution and during training, it selects examples such a way that the overall distribution of the training data would fit the desired one. This new distribution is a linear combination of the original and a uniform one, $\lambda$ and $1-\lambda$ being the respective coefficients.

These resampling methods are easy to use; however, we had to adapt them to work in a multi-task setup. To upsample, we created clusters, which had the same label pair, and resampled so that each group would have the same amount of training data. Although this adaptation does not ensure the individual tasks having balanced data, in practice, it works quite well, as shown in section \ref{ssec:elderly}. A similar modification can be applied when using probabilistic sampling in a multi-task setting. First, we generate the desired distribution for each task. Then during training, we select a label pair that would fit the distribution and use a training instance that has those labels.

\subsection{Low-frequency features for mask sub-challenge}
\label{ssec:low-freq-feats}
\begin{figure}
    \centering
    \includegraphics[trim={0 0 1.5cm 0},clip, height=7cm]{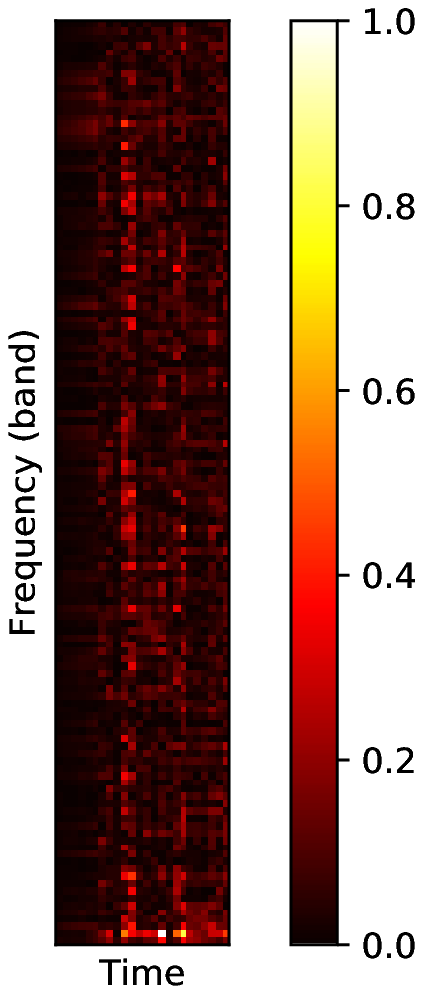}
    \includegraphics[trim={0.5cm 0 1.5cm 0},clip,height=7cm]{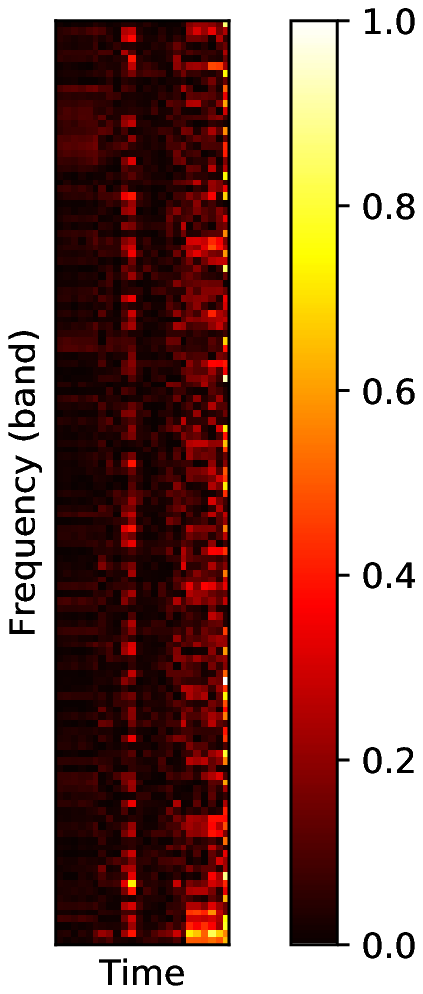}
    %\hfill
    \includegraphics[trim={0.5cm 0 1.5cm 0},clip,height=7cm]{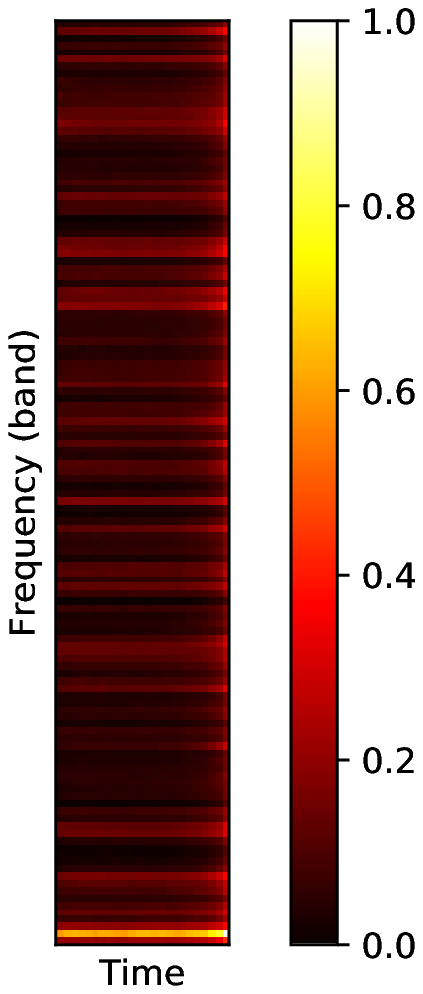}
    \includegraphics[trim={0.5cm 0 0 0},clip,height=7cm]{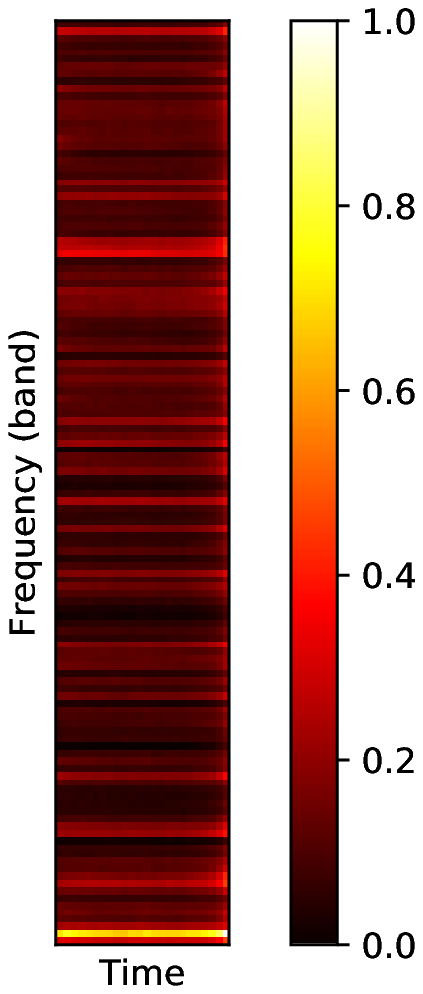}
    \caption{The figure shows the gradient magnitudes of DNN outputs with respect to the input spectral features on the Mask sub-challenge. The left two images show the gradients of the two random models for a single audio file and the right two images show the gradients of the same two models averaged over all training files. A bright yellow shade represents the largest gradient magnitudes seen for the lowest frequencies.}
    \label{fig:low-freq-grads}
    \vspace{-0.5cm}
\end{figure}

For the mask sub-challenge, we hypothesize that wearing a mask changes the resonance conditions in the vocal tract, as the mask might reflect some of the frequencies to the tract \cite{phonation_speech,Mel_phonation}. To test this hypothesis, we look at the output gradients w.r.t. the inputs and plot them per input frequency bands in Figure \ref{fig:low-freq-grads}. We notice that end-to-end models have large gradients for the ten lowest frequencies. Considering this observation, we compute low-frequency information-based features. Specifically, we extract Mel-spectrogram features for 200 filter-banks and then use the ten lowest filter-banks as input features, which is referred to as \textbf{lowest-10-features}. 

As a pre-processing step for extracting these features, we also examine enhancing the lower frequencies by manipulating the input audio. We apply low-frequency enhancing schemes like preemphasing the audio (with filter coefficients $h= 1$ and passing through a fifth-order low-pass butterworth filter whose cutoff frequency is $400~Hz$ \cite{oppenheimBook}, denoted as \textbf{preemphasis+butterworth}. These schemes can allow the Mel-spectrogram to better represent the relevant information for this task, which is dependent on the low-frequency bands.

\section{Experiments and results}

\subsection{Experimental setup}
\label{ssec:models}
%exact details, net structure, optimizer, framework etc.
For the elderly and mask sub-challenges, we extract Mel-spectrograms from the audio files as inputs in a similar fashion to the auDeep \cite{Freitag2017AuDeep} pipeline. Instead, for the breathing sub-challenge, the raw audio data directly input as the raw audio leads to better results than Mel-spectrograms. 

In our experiments, we use two different end-to-end systems. For the elderly and mask sub-challenges,  the spectral input is first processed by a 1D convolutional layer with 100 neurons and then a recurrent layer, containing 100 LSTM cells, accumulates the outputs of the filters. We pass the outputs of the recurrent layer to a feedforward layer (100 rectified-linear units) and then apply a classification layer. In the multi-task experiments, we split the structure after the LSTM layer passing the recurrent layer output to a unique set of hidden and output layers for both tasks. For the breathing sub-challenge, we opted for the same structure as the best baseline system, for details see~\cite{compare2020}. For training, we employed the Keras framework with TensorFlow as the backend. 

For all tasks, we use ensemble learning. For the mask sub-challenge, we obtained the best results using 50 models, while for the other tasks, ten models were enough to reach the peak performance. After training the individual models, we averaged their output to create the final predictions.

%talk about the ?-marks in tables
For evaluation on the test set, we train our models on the combined training and development set. We note that the ComParE challenge restricts the number of submissions per team and task to five evaluations on the test set. As the competition is ongoing, we only used a few of the available submissions to check the best systems so far. The limitation implies that we can not test all of our methods. In the result tables, we use the question mark (?) to indicate the solutions not yet evaluated on the test data.

\subsection{Breathing sub-challenge}
On this task, we used the end-to-end baseline system for further development as this system performs quite well on this task \cite{compare2020}. %However, there are a few speakers for which the official evaluation metric, the Pearson correlation, reports a much lower correlation value than for the others. To normalize the impact of these outliers, we also evaluate using the correlations per person averaged together (Average corr).
%We feel that this metric better represents the system performance, and shows these per speaker correlations to highlight the effect that a few outliers can have on the overall correlation. 
However, it faces the issue of mismatch on the scale of the end-to-end predictions and output labels. To alleviate this mismatch, we apply a multi-loss scheme using MSE based loss to regularize the baseline correlation loss.

\begin{table}[!t]
    \centering
        \caption{The table presents the breathing task's Pearson correlation scores for single end-to-end models on the development set (Dev). As we trained ten different models, we present the average and the best result.}
    \begin{tabular}{c|c|c}\hline
        {\bf System} & \multicolumn{2}{c}{{\bf Dev}}\\ \hline
        & Avg. per DNN& Best DNN\\
        \hline
        E2E-corr & .506 & .514\\
        E2E-MSE & .467 & .481 \\
        E2E-corr+MSE & .497 & .521\\ \hline

    \end{tabular}
    \label{tab:breath_ens_vs_single}
    
\end{table}
In table~\ref{tab:breath_ens_vs_single}, we compare between the single-loss versus multi-loss strategies. The single-loss models use either Pearson's correlation (corr) or the MSE. The multi-loss strategy combines the two losses (corr+MSE) with a regularization weight of 0.1. Correlation-based E2E (E2E-corr) performs best on when averaging the correlation values of ten randomly-initialized DNNs. The best result corresponds to the corr+MSE based E2E model; however, the averaged results are lower and suggest that this value is unreliable. We suspect further tuning of the regularization weight is required and we hope to complete this analysis as part of our future work.

% \begin{table}[!t]
%     \centering
%         \caption{The table presents correlation values achieved on the breathing task by using different loss functions to train the ensemble E2E models.}
%     \begin{tabular}{c|c|c|c| c}\hline
%         {\bf System (loss function)} & \multicolumn{3}{c|}{{\bf Dev}} & {\bf Test}  \\ \hline
%         & Corr & Avg. corr& MSE & Corr\\
%         \hline
%         Baseline (E2E) \cite{compare2020} & .507 & .629 &  1.682 & .731 \\
%         \hline
%         E2E-corr & \textbf{.523} & .644 & .896 & \textbf{.759}\\
%         E2E-MSE & .480 & .593 & .028 & ? \\
%         E2E-corr+MSE & .514 & .632 & .180 & .751\\ \hline
%     \end{tabular}
%     \label{tab:breath}
%     \vspace{-0.5cm}
% \end{table}
\begin{table}[!t]
    \centering
        \caption{The table presents the ensemble E2E model's performance on the breathing task for different loss functions.}
    \begin{tabular}{c|c|c| c}\hline
        {\bf System (loss function)} & \multicolumn{2}{c|}{{\bf Dev}} & {\bf Test}  \\ \hline
        & Corr & MSE & Corr\\
        \hline
        Baseline (E2E) \cite{compare2020} & .507 &  1.682 & .731 \\
        \hline
        E2E-corr & \textbf{.523} & .896 & \textbf{.759}\\
        E2E-MSE & .480 & .028 & ? \\
        E2E-corr+MSE & .514 & .180 & .751\\ \hline
    \end{tabular}
    \label{tab:breath}
    \vspace{-0.5cm}
\end{table}
In table~\ref{tab:breath}, we present the results of 10-model ensembles to compare with the baseline performance. Even though the baseline system produced high correlation values, it had the highest MSE value. Combining the predictions of 10 models (corr) reduced the MSE significantly and outperformed the baseline results. Using the MSE as loss function performed the worst but, naturally produced the lowest MSE. Lastly, we can see that using the multi-loss ensemble of E2E model (E2E-corr+MSE) drops in comparison to the ensemble of  E2E-corr because of the MSE regularization. However, in terms of MSE, it is much better. On the evaluation set, E2E-corr+MSE was also slightly worse than the E2E-corr ensemble in terms of overall correlation. Nevertheless, our ensemble of E2E corr outperforms the baseline result and shows an absolute improvement of 2.6 correlation points over the baseline result. %In terms of average correlation, our E2E-corr. %and E2E-corr.+MSE ensembles report higher correlation than the official metric as it is less sensitive to outliers.

\subsection{Elderly sub-challenge}
\label{ssec:elderly}
The elderly task presents a prediction problem with class imbalance. For valence prediction, 44 out of the 87 stories have a medium-valence label. Upon inspecting some of our initial models, we observe that the output prediction favours the over-represented classes. To cope with these issues, we apply the resampling methods described in section~\ref{sec:resample}.

Table \ref{tab:elderly} presents the ensemble E2E models evaluated on the elderly sub-challenge. Applying the sampling techniques improves the performance significantly in each case. For arousal, upsampling showed more benefit than probabilistic sampling ( Table\ref{tab:elderly}). In contrast, probabilistic sampling with $\lambda=0.6$ was very beneficial for the valence sub-task. The multi-task models consistently outperformed single task ones. For the two best systems, we also checked the performance of the individual DNNs and saw that ensemble learning is essential for good performance. Upsampling for a single multi-task DNNs on average yielded 38.4\%/36.4\% (A/V), with probabilistic sampling we got 36.6\%/38.5\%. 

Unfortunately, the test results are below the official baseline. The considerable difference between scores on the development and test data suggest that our model overfits when training a train+development set system for evaluation.%, and this warrants the use of regularization techniques in the future.
\begin{table}[!t]
    \centering
        \caption{UAR values for predicting Arousal (A) and Valence (V) levels on the elderly sub-challenge. E2E systems combined ten DNNs to produce the predictions.}
    \begin{tabular}{c|c|c}\hline
        {\bf System} & {\bf Dev (A/V)} & {\bf Test (A/V)}  \\
        \hline
        Baseline (linguistic) \cite{compare2020} & 40.6/\textbf{49.2} & 44.0/\textbf{49.0} \\
        \hline
        Baseline(acoustic) \cite{compare2020} & 35.0/31.6 & \textbf{50.4}/40.3\\

        E2E (single task) & 35.0/39.7  & ?\\
        E2E (multitask) & 39.5/39.7 & ?\\
        E2E (single task + upsampl.) &  39.8/41.5 & ?\\
        E2E (multitask+ upsampl.) & \textbf{42.9}/42.4  & 38.0/39.5\\
        E2E (single task + prob. sampl.) & 35.6/39.6  & ?\\
        E2E (multitask+ prob. sampl.) & 40.0/\textbf{45.5} & 45.8/34.8\\ \hline
    \end{tabular}
    \label{tab:elderly}
    % \vspace{-0.5cm}
\end{table}
%talk about the baselines and how the organizers have cheated
We also suspect that there is a significant mismatch between the dev and test data in this sub-challenge. Strong evidence for this can be found in the baseline paper~\cite{compare2020}. In the baseline paper, we can see that the test performance does not correlate with the scores achieved on the development set. The official acoustic baseline model (DeepSpectrum+SVM) produces almost the worst results on the development set, and the difference between its development and test scores is large. This observation suggests that parametric tuning with the development data might not be the best model for the evaluation set.

\subsection{Mask sub-challenge}
Training a 50 model ensemble, we saw that their averaged prediction significantly outperformed our single E2E model and the individual baseline system (auDeep-fused). Our individual E2E models achieved 66.0\%UAR on average, but their combination reached 68.0\% (E2E). The best individual baseline uses auDeep-based features in an SVM system. Our E2E ensemble outperformed this model both on the development and test set, as shown in table~\ref{tab:mask}. Though, our ensemble E2E model is outperformed by the fusion of the best baseline models, which is an SVM based on auDeep-fused, Bag-of-audio-word, OpenSmile and DeepSpectrum features \cite{compare2020}.

Earlier in section \ref{ssec:low-freq-feats}, we had observed that lower-frequency bands of the audio hold important information for the mask sub-challenge. Based on this observation, we applied preemphasis+butteworth to input audio and then extracted the lowest-10-features to build an E2E ensemble (E2E lowest-10-features). This ensemble model outperformed the E2E-ensemble built with only preprocessed input audio (E2E preemphasis+butterworth) but was worse than our vanilla ensemble. 

Combining the regular ensemble (E2E) and E2E lowest-10-features fared better, resulting in a slight improvement over the vanilla ensemble. We combined this model with predictions from SVMs trained on bag-of-audio-word features (BoAW-fused) and DeepSpectrum-resnet50 features (E2E+lowest-10-feats+baseline) via soft voting with equal weights. The combined achieved our best result on the development set and improved the fusion-based baseline by  3.8\% UAR.

\section{Future work}
\begin{table}[!t]
    \centering
        \caption{UAR values on the Mask Sub-Challenge. The E2E solutions fused 50 models to get the final output.}
    \begin{tabular}{c|c|c} \hline
        {\bf System} & {\bf Dev} & {\bf Test}  \\
        \hline
        auDeep-fused (baseline) \cite{compare2020}& 64.4 & 66.6\\
        Fusion of the bests (baseline) \cite{compare2020}& -- & 71.8 \\
        \hline
        E2E & 68.0 & 69.9\\
        %E2E lowest-2-bands& 60.3& ?\\
        E2E preemphasis+butterworth & 59.3 & ?\\
        E2E lowest-10-features & 62.9 & ? \\
        E2E + E2E lowest-10-features & 68.6 & ? \\
        E2E + E2E lowest-10-features + baseline & \textbf{70.2} & \textbf{75.6} \\ \hline
    \end{tabular}
    \label{tab:mask}
    % \vspace{-0.5cm}
\end{table}
For the breathing sub-challenge, we observe that regularizing with MSE can help alleviate the mismatch of scales between the output and the labels. However, it still lacks in performance in comparison to the regular E2E model. To study this effect, we explore other regularization schemes to obtain a better balance between mismatch issues and performance.

On the elderly sub-challenge, our current system overfits on the training and development data and obtains a poor performance on the test set. We investigate this effect further and apply regularization schemes to reduce the overfitting. Another thing that limits our current system is that it is trained to classify short segments of the stories and then the decisions made for these fragments are merged with a soft voting method. Instead, we could concatenate the audio files of the same stories and directly classify them, as our E2E architecture allows us to use arbitrary long inputs.

For the mask sub-challenge, we currently combine predictions from separate models for the vanilla and lowest-10-features E2E scenarios. In contrast with this late fusion, we explore early and intermediate fusion of features to better exploit the information present in these spectrograms. Our lowest-10-features na\"ively extracts the ten lowest frequency bands to use as features for the E2E model. Instead, we also develop specialized low-frequency features to aid better learning by the E2E model. In the mask sub-challenge, we observe that combining our E2E ensembles with baseline results achieves the best result. We plan to explore similar combinations for both breathing and elderly sub-challenges.

\section{Conclusions}
We presented Aalto's E2E ensemble solution for the three different INTERSPEECH 2020 ComParE tasks. In our study, the ensemble E2E models achieved better performance than individuals E2E models on average. On the ComParE 2020 tasks, we also proposed task-specific modifications for the underlying E2E models. We studied modifications based on multi-task learning, re-sampling training data for multi-task scenarios, and feature engineering based on the initial E2E ensemble models. Our best models showed absolute improvements upon the competitive baselines for the breathing and mask sub-challenges by 2.8\% and 3.8\%, respectively. Overall, our paper showcased the benefits of using an ensemble of E2E models and task-specific modifications for computational paralinguistic tasks.

%\section{Acknowledgements}
%Tam\'as Gr\'osz was supported by the Kone Foundation. 
%Sudarsana Kadiri was supported by the Academy of Finland (project no. 312490).
\section{Acknowledgements}
We thank Antonia Hamilton and Alexis Macintyre for granting us access to a subset of the UCL Speech BreathMonitoring (UCL-SBM) database used in the Breathing Sub-Challenge.
This work was supported by the Academy of Finland (grants 312490, 329267) and the Kone Foundation.  Aalto ScienceIT provided the computational resources.

\bibliographystyle{IEEEtran}
\bibliography{compare}
\end{document}

%% file: abstract.tex
End-to-end neural network models (E2E) have shown significant performance benefits on different INTERSPEECH ComParE tasks. Prior work has applied either a single instance of an E2E model for a task or the same E2E architecture for different tasks. 
However, applying a single model is unstable or using the same architecture under-utilizes task-specific information. On ComParE 2020 tasks, we investigate applying an ensemble of E2E models for robust performance and developing task-specific modifications for each task. ComParE 2020 introduces three sub-challenges: the breathing sub-challenge to predict the output of a respiratory belt worn by a patient while speaking, the elderly sub-challenge to estimate the elderly speaker's arousal and valence levels and the mask sub-challenge to classify if the speaker is wearing a mask or not. On each of these tasks, an ensemble outperforms the single E2E model. On the breathing sub-challenge, we study the impact of multi-loss strategies on task performance. On the elderly sub-challenge, predicting the valence and arousal levels prompts us to investigate multi-task training and implement data sampling strategies to handle class imbalance. On the mask sub-challenge, using an E2E system without feature engineering is competitive to feature-engineered baselines and provides substantial gains when combined with feature-engineered baselines.

%% file: introduction.tex
\section{Introduction}
INTERSPEECH's Computational Paralinguistics Challenges (ComParE) have regularly introduced the speech community to new exciting challenges since 2009\footnote{\url{http://www.compare.openaudio.eu/}}. These challenges setup as prediction tasks focus on extracting important speaker-related information from the audio signal. In more than a decade of ComParE, researchers have come up with innovative solutions to these challenges.

These efforts can broadly be divided into two categories: feature-engineering based approaches \cite{Schmitt2016BOAW, Eyben2016GeMAPS, Freitag2017AuDeep, Amiriparian2017DeepSpect, Montacie2018, Gosztolya2018, Gosztolya2019, Wu2019, Carbonneau2020Feature} and deep learning-based end-to-end approaches \cite{0001SSA18,Zhao2018DeepSpect,ElsnerLRMI19,DBLP:journals/corr/abs-1802-01115,Fritsch2020RawCNN}. Feature-engineering approaches have concentrated on extracting task-specific features to be utilized by classifiers for prediction. On the other hand, end-to-end approaches have focused applying complex neural network architectures to bypass feature engineering. There might not be a clear winner between these two approaches~\cite{0001SSA18}, but combining these two approaches has emerged as a trend. Prior work has used Deep Neural Networks (DNN) only to extract useful features automatically. These features are used to train a simple classifier afterwards. Two such systems, AuDeep~\cite{Freitag2017AuDeep}, and DeepSpectrum~\cite{Amiriparian2017DeepSpect} are already part of this year's baseline method. %Both of them have a similar concept; they extract the spectrum of the audio data and use a DNN to generate some latent feature vector.  The main difference between them is that AuDeep uses a recurrent Autoencoder, and DeepSpectrum utilizes a pre-trained Convolutional Neural Network (CNN).

For end-to-end (E2E) approaches, prior work has concentrated on applying single-model systems for the prediction tasks \cite{GosztolyaBGT17,0001SSA18,ElsnerLRMI19,DBLP:journals/corr/abs-1802-01115}. A single neural network, which is a non-linear model, can have high-variance and, thus, produce unstable results. On the other hand, the prior work has concentrated on applying the same architecture across different tasks. Using the one-size-fits-all policy for different ComParE tasks ignores task-specific requirements that can be exploited in model design~\cite{0001SSA18}. Hence, we study the application of E2E models to obtain a more robust but task-specific solution. We build ensemble-based E2E systems to obtain robust results across different ComParE 2020 tasks. Utilizing multiple models instead of one shows better performance across all the tasks. Besides, we also study task-specific requirements and explore incorporating them into the E2E solution.% to help with learning a better prediction system. 

ComParE 2020 poses three new challenges \cite{compare2020} to the community: 1) the breathing sub-challenge to predict the output signal of a respiratory belt worn by the speaker, 2) the elderly sub-challenge to classify the arousal (A) and valence (V) level of elderly speakers and 3) the mask sub-challenge to predict if the speaker wears a mask or not while they speak.

For the breathing sub-challenge, the E2E baseline system optimizes Pearson's coefficient, which is also the task metric, to solve a regression problem. However, E2E predictions do not match the scale of the ground truth. To alleviate this scale issue, we study multi-loss strategies for our E2E model, where we optimize Pearson's coefficient along with mean-squared-error. The elderly sub-challenge entails learning two closely related tasks (arousal and valence level prediction), so as a natural choice, we explore multi-task learning. The sub-challenge also faces the issue of imbalanced class data and we apply sampling schemes to augment the data to reduce the class imbalance. 
In the mask sub-challenge, our single E2E model performs better than the best baseline result. On further investigation, we analyze the trained models to understand which frequency bands hold the largest importance. Our findings lead us to create low-frequency band features. A fusion of the baseline with our E2E models, including these features, results in a substantial performance gain.

We also include our plans to explore and expand our presented experiments. We hope to include these results in an extended version of this article.